\documentclass[aps,prl,notitlepage,superscriptaddress,showpacs,twocolumn]{revtex4-2}
\usepackage{graphicx,subfigure,epsfig}
 \usepackage{array,multirow}
\usepackage{dcolumn}
\usepackage{amssymb,amsmath,amsfonts,mathrsfs}
\usepackage{array}
\usepackage{times,setspace}
\usepackage{latexsym}
\usepackage{float,flafter,bm,bbm}
\usepackage{epstopdf,color,multirow}
\usepackage[colorlinks,linkcolor=blue,anchorcolor=blue,urlcolor=blue,citecolor=blue]{hyperref}
\usepackage{footnote}
\usepackage{booktabs}
\hypersetup{
    colorlinks=true,
    linkcolor=blue,
    filecolor=magenta,
    urlcolor=blue,
}

\begin{document}
\title{The Ginzburg-Landau theory of flat band superconductors with quantum metric}
\author{Shuai A. Chen}
\email{chsh@ust.hk}

\affiliation{Department of Physics, Hong Kong University of Science and Technology,
Clear Water Bay, Hong Kong, China}
\author{K. T. Law}
\email{phlaw@ust.hk}
\affiliation{Department of Physics, Hong Kong University of Science and Technology,
Clear Water Bay, Hong Kong, China}
\date{\today}

\begin{abstract}
Recent experimental study unveiled highly unconventional phenomena in the superconducting twisted bilayer graphene (TBG) with ultra flat bands, which cannot be described by the conventional BCS theory. For example, given the small Fermi velocity of the flat bands, the superconducting coherence length predicted by BCS theory is more than 20 times shorter than the measured values. A new theory is needed to understand many of the unconventional properties of flat band superconductors. In this work, we  establish a Ginzburg-Landau (GL) theory from a microscopic flat band Hamiltonian. The GL theory shows how the properties of the physical quantities such as the critical temperature, the superconducting coherence length, the upper critical field and the superfluid density are governed by the quantum metric of the Bloch states. One key conclusion is that the superconducting coherence length is not determined by the Fermi velocity but by the size of the optimally localized Wannier functions which is limited by quantum metric. Applying the theory to TBG, we calculated the superconducting coherence length and the upper critical fields. The results match the experimental ones well without fine tuning of parameters. The established GL theory provides a new and general theoretical framework for understanding flat band superconductors with quantum metric.
\end{abstract}

\maketitle

\emph{\color{blue}Introduction.---} 
Our understanding of quantum states of matter has been greatly deepened by the study of the geometric properties of Bloch wavefunctions in crystals. More specifically, the imaginary and real parts of the quantum geometric tensor of Bloch wavefunctions, which are the Berry curvature and the quantum metric respectively, greatly influence the properties of the quantum states \cite{1980CMaPh76289P,berry1989quantum}.
The Berry curvature arises from the phase difference between two neighboring Bloch states and characterizes the band topology of states such as the quantum Hall and the Chern insulating states \cite{prlIQH,1982Thoulessprl,1984Berry,1994JMP355373B,TIRMP,TISCRMP,berry1989quantum,2023arXiv230302180B,2017AnPhy.377..345C}. On the other hand, the quantum metric measures the distance between two adjacent Bloch states \cite{1990AAgeometry,2011EPJB79121R}. It describes the wave function extension and quantifies the level of obstruction of an exponentially localized Wannier basis \cite{PhysRevB.56.12847}. The quantum metric property is important for the formation of fractional quantum Hall and fractional Chern insulating states \cite{PhysRevLett.107.116801,PhysRevB.33.2481,2002LNP59598F,2013CRPhy14816P,PhysRevB.88.115117,PhysRevB.90.165139,2022arXiv221013487W}. More recent studies have shown the fundamental roles of quantum metric in various physical phenomena, including quantum transport and electromagnetic responses \cite{PhysRevB.87.245103,PhysRevLett.115.166802,PhysRevLett112166601,PhysRevB.94.134423,PhysRevResearch.3.L042018,PhysRevB.104.L100501,PhysRevLett.126.156602,2021NatPh..18..290A,2022PhRvB.105h5154M}, superfluidity and superconductivity in flat bands \cite{2015NatCo68944P,PhysRevB.95.024515,PhysRevA.97.033625,PhysRevB.98.220511,2020arXiv200716205J,PhysRevB.102.201112,PhysRevA.101.053631,PhysRevLett.117.045303,PhysRevLett.127.170404,PhysRevLett.128.087002,PhysRevB.106.014518,PhysRevB.105.L140506,PhysRevB.106.104514,2022arXiv220900007H,2022arXiv220402994H}, and quantum phase transitions \cite{2006PhRvB..74w5111T,PhysRevLett99095701,PhysRevLett99100603,2021PNAS11806744V,2023PNAS..12017816M}. In particular, the effect of quantum metric on the properties of moir\'e materials has attracted much attention in recent years \cite{2018Natur.556...80C,2018Natur55643C,2019Natur.574..653L,PhysRevX.9.031049,PhysRevLett.123.237002,PhysRevB.101.060505,PhysRevResearch.2.023237,2020PhRvL124p7002X,PhysRevResearch4013164,PhysRevB.104.115160,PhysRevResearch.4.013209,2021arXiv211100807T,PhysRevLett.127.246403,PhysRevLett.128.176403}.

The quantum metric effect on superconductivity in flat band systems with vanishing Fermi velocity $v_F$ is particularly interesting. On one hand, according to the BCS theory, a large pairing gap $\Delta$ and a high critical temperature $T_c$ are expected due to the large density of states of flat bands. Moreover, the relation $\xi = \frac{\hbar v_{F}}{\Delta}$ seemingly implies a vanishing short coherence length $\xi$ such that electrons are tightly bound to form Cooper pairs.  These BCS relations point to a very robust superconducting state in flat band superconductors. On the other hand, the diverging effective mass implies a vanishing superfluid weight $\mathsf{D}_{s}$ as $\mathsf{D}_{s}\!\propto\!1/m^{*}\!\!\rightarrow0$. This implies the absence of supercurrents and the Meissner effect which define superconductivity. Recently, Peotta and {T{\"o}rm{\"a}} \cite{2015NatCo68944P} shed light on the problem by pointing out that a supercurrent is indeed achievable and the superfluid weight is proportional to the quantum metric for the flat bands \cite{2015NatCo68944P,PhysRevB.95.024515,PhysRevB.106.014518}.  

Very recently, the superconducting properties of twisted bilayer graphene (TBG) with an extremely low Fermi velocity of $v_F \approx 1,000$m/s were studied experimentally. It was shown that many of the superconducting properties deviate greatly from the conventional BCS predictions \cite{2023Natur.614..440T}. For example, the coherence length is estimated to be around $2.6$nm according to the BCS relation $\xi = \frac{\hbar v_{F}}{\Delta}$, which is much shorter than the estimated value of 55nm (at optimal doping) according to the upper critical measurements. Due to the large effective mass of the electrons, it is also expected that the superfluid stiffness, which is proportional $\frac{1}{m^{*}}$, is low.  The  Berezinskii-Kosterlitz-Thouless (BKT) transition temperature is estimated to be about 0.05K, which is much lower than the measured $T_c = 2.2$K at optimal doping. In short, BCS relations which connect physical quantities with $v_F$ or $m^{*}$ have failed to provide a proper description of superconductivity in TBG. A new theory is therefore needed to understand flat band superconductivity.

\begin{table}[]
\caption{Comparison between the BCS theory and the GL theory of flat band superconductors. The results for the superfluid weight $\mathsf{D}_s^{ab}(T)$ at temperature $T$, the superconducting transition temperature $T_c$, the superconducting coherence length $\xi$, 
and the upper critical field $H_{c2}$ are summarized. $T_c=T_{\text{BKT}}$ for the flat band superconductor. The quantum metric $g_{ab}$ is defined in Eq.~\eqref{eq:gab} and $\bar{g}_{ab}$ is averaged over the first Brillouin zone as in Eq.~(\ref{eq:aver_gamma}). 
$\mathcal A_\mathrm{uc}$ is the area of a unit cell, $\chi_2^{ab}$ is defined in Eq.~\eqref{eq:chiab}, and $\Phi_0=h/2e$ is the flux quantum.
The superfluid density for BCS theory is denoted as $n_s$.
The experimental values are adopted from Ref.~\onlinecite{2023Natur.614..440T}. An effective attractive interaction strength of $U=0.6$meV is used for calculating the $T_{c}$ for TBG. 
}
\label{tab:summary}
\begin{tabular}{ccccc}
\toprule
                  & BCS      & Flat band & TBG(Exp.) & Theory \\ 
\midrule
$\mathsf{D}_s^{ab}$ & $\delta_{ab}\frac{n_{s}(T)}{m}$       &  $\mathcal A_\mathrm{uc}^{-1}\chi_2^{ab}(T)$ &    \\
$T_c$               & $ 1.7^{-1}U\Delta_{0}$        & $\frac{\pi}{8}\sqrt{\mathrm{det}\mathsf D_s^{ab}}$ &    $2.2$K  & $1.66$K  \\
$\xi$               & $\frac{\hbar v_{F}}{U\Delta_0}$     &  $\sqrt{\frac{T_\mathrm{MF}}{\vert T- T_\mathrm{MF}\vert}  }[\mathrm{det}(\bar{g}_{ab})]^{\frac{1}{4}}$  &  $55$nm &$ 30$nm  \\
$H_{c2}$            &  $2\pi (\frac{T_{c}}{v_{F}})^{2}$      &   $\frac{\vert T-T_\mathrm{MF}\vert}{T_\mathrm{MF}}\frac{\Phi_0}{2\pi\sqrt{\det(\bar{g}_{ab})}}$  &   $0.10$T &  $ 0.27$T     \\
\bottomrule
\end{tabular}
\vspace{-10pt}
\end{table}

\begin{figure}[th]
\centering \includegraphics[scale=0.68]{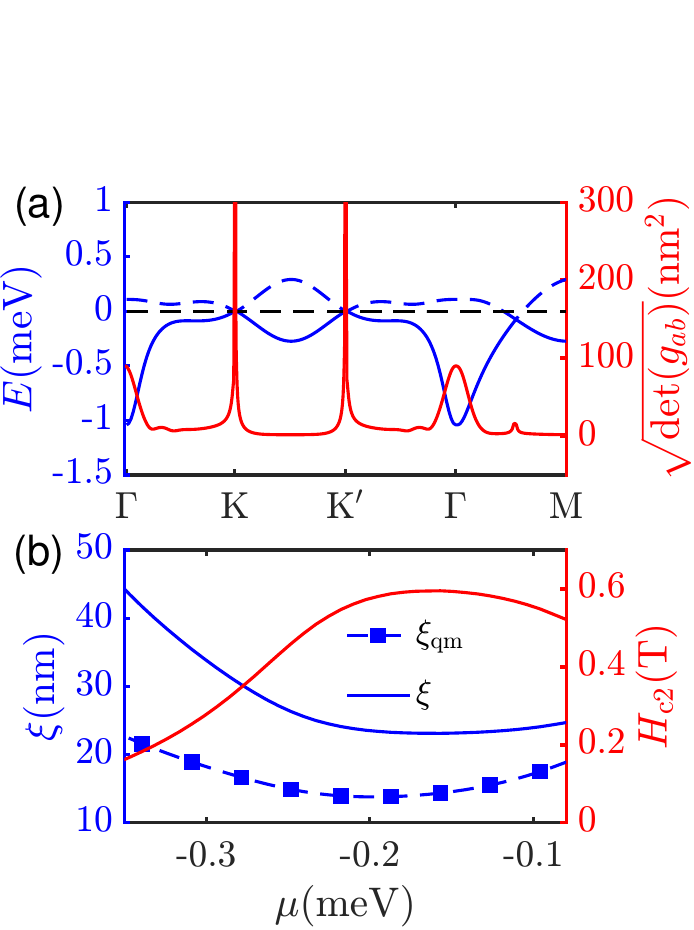} 
\caption{
(a) The band structure and the distribution of the quantum metric $\sqrt{\mathrm{det}(g_{ab})}$ of the Bloch states of the moir\'e BZ of TBG. The quantum metric is plotted for the partially filled hole band (blue solid line). The quantum metric diverges at the two Dirac points at $\mathrm K$ and $\mathrm K^\prime$. For simplicity, only states originated from the $\mathrm K$ valley of the original BZ are displayed. (b) The superconducting coherence length $\xi$ as a function of chemical potential $\mu$ at $0.2$K. Here $\xi_\mathrm{qm}$ is the quantum metric contribution to $\xi$. As doping moves away from $\mu=-0.1$meV, $\mu$ decreases, but increases again around $\mu=-0.2$meV. An effective interaction strength $U=0.6$meV is assumed in the calculations.
}
\label{Fig:TBG} 
\end{figure}

In this work, we develop the Ginzburg-Landau (GL) theory of flat band superconductors by incorporating the quantum geometric properties of the Bloch electrons. 
Besides reproducing previous results concerning the BKT transition temperature \cite{2015NatCo68944P} and the superfluid weight \cite{2015NatCo68944P}, the GL theory allows us to determine the coherence length and the upper critical field and their dependence on the quantum metric. The results are summarized in Table.~\ref{tab:summary}. Applying our theory to TBG with a small Fermi velocity, we estimated the coherence length, and the upper critical field which match the experimental measurements well without the fine tuning of parameters as shown in Table.~\ref{tab:summary} and Fig.~\ref{Fig:TBG}. A striking result concerning $\xi$ in the flat band limit is that it is independent of interaction strength at zero temperature and is purely determined by the quantum metric effect (See also Eq.~(\ref{eq:cohlen})). Contrary to the conventional understanding that a stronger interaction will bind electrons closer together to reduce the Cooper pair size (which is measured by~$\xi$), the quantum metric limits the size of the Cooper pairs. 
The Cooper pair size cannot be smaller than the size of optimally localized Wannier basis \cite{PhysRevB.56.12847} constructed by the Bloch states. In the case of TBG, the quantum metric (together with a small band dispersion) limits $\xi$ to be tens of nanometers as observed in the experiment \cite{2023Natur.614..440T}. 

In the following, we first derive the GL free energy which incorporates the quantum metric effects of Bloch electrons. Second, the superfluid weight, the upper critical field and the superconducting coherence length are derived. Finally, we apply the GL theory to explain the unconventional behavior of superconducting TBG.

\emph{\color{blue} The Ginzburg-Landau Free Energy.---} 
We start with a Hamiltonian $H=H_{0}+H_{\mathrm{int}}$. It is assumed that $H_{0}$ is the non-interacting part of a multi-band Hamiltonian which possesses isolated flat bands at the Fermi energy. 
In general, the flat bands have Bloch states of the form 
$e^{-i\mathbf{q}\cdot\mathbf{r}}u_{\mathbf{q}\xi}$, where $\mathbf{q}$ denotes the crystal momentum and $\xi$ is the flavor index. For a multi-orbital system, the Bloch wavefunctions $u_{\mathbf{q}\xi}$ possess multiple orbital components such that $u_{\mathbf{q}\xi} = \sum_{\alpha} u_{\mathbf{q}\xi}(\alpha) $ where $\alpha$ is the orbital index. Even though the bands are completely flat, there can be nontrivial quantum geometry effects encoded by $u_{\mathbf{q}\xi}$. 
The occurrence of a superconducting phase is associated with an effective attractive interaction which can be written as:
\begin{equation}
H_{\mathrm{int}}=-U\int d\mathbf r ~ a_{+}^{\dagger}(\mathbf{r})a_{-}^{\dagger}(\mathbf{r})a_{-}(\mathbf{r})a_{+}(\mathbf{r})~,\label{eq:Hint}
\end{equation}
where $a_{\xi}(\mathbf{r})$ is an electron annihilation operator carrying two flavors $\xi=\pm$. To understand the role of quantum geometry with interactions, we project the electron operators $a_{\xi}(\mathbf{r})$ to the Bloch states of the relevant flat bands near the Fermi energy (see Supplemental Material, Section I (SM-I)~\cite{SM}) such that:
\begin{equation}
a_{\xi}(\mathbf{r})\rightarrow\frac{1}{\sqrt{N}}\sum_{\mathbf{q}}\sum_{\alpha}e^{i\mathbf{q}\cdot\mathbf{r}}u_{\mathbf{q}\xi}^{*}(\alpha)c_{\mathbf{q}\xi}~,\label{eq:projectiongk}
\end{equation}
where $c_{\mathbf{q}\xi}$ annihilates an electron with momentum $\mathbf{q}$, 
and the sum of $\mathbf{q}$ is over the first Brillouin zone (BZ). The expansion in Eq.~(\ref{eq:projectiongk}) projects out other bands which are far away in energy from the relevant flat bands. 
We proceed with the Hubbard-Stratonovich transformation by introducing
a bosonic field $\Delta(\mathbf{r})$, 
\begin{align}
\Delta(\mathbf{r}) & =a_{-}(\mathbf{r})a_{+}(\mathbf{r})~,
\end{align}
Then, the Lagrangian density $\mathcal{L}$ is obtained through the path integral approach (SM-I~\cite{SM}) such that
\begin{align}
\mathcal{L}= & (-i\omega-\mu)(\bar{c}_{\mathbf{k},+}c_{\mathbf{k},+}+\bar{c}_{\mathbf{k},-}c_{\mathbf{k},-})\nonumber \\
 & -U\sum_{\mathbf{q}}[\Gamma(\mathbf{q},\mathbf{k})\Delta(\mathbf{k})\bar{c}_{\mathbf{q}+\mathbf{\frac{k}{2}},+}\bar{c}_{\mathbf{-q}+\frac{\mathbf{k}}{2},-}+h.c.],
\end{align}
where $c_{\mathbf{q},\xi}$ denotes the Grassmann fields, $\mu$ is
the chemical potential and  $\Delta(\mathbf{k})\equiv \sum_{\mathbf{r}}\Delta(\mathbf{r})e^{i\mathbf{k}\cdot\mathbf{r}}$ is the Fourier component of the bosonic field $\Delta(\mathbf{r})$. 
The projection in Eq.~\eqref{eq:projectiongk} introduces the form factor  $\Gamma(\mathbf{q},\mathbf{k})$ which
modifies the coupling constant $U$. The form factor is defined as $\Gamma(\mathbf{q},\mathbf{k})\equiv\sum_{\alpha}u_{\mathbf{\mathbf{-q}+\mathbf{k}/2},+}(\alpha)u_{\mathbf{\mathbf{q}+\mathbf{k}/2},-}(\mathbf{\alpha})$,
and it plays a crucial role in the context of superconductivity.
Formally, the GL free energy $F[\Delta]$ is obtained by integrating out the fermion fields
at a finite temperature $T$ such that, 
\begin{equation}
\!\!F[\Delta]=\sum_{\mathbf{k}}\! U\bar{\Delta}(\mathbf{k})\Delta(\mathbf{k})-T\ln\int\mathcal{D}[c,\bar{c}]e^{-\int_{0}^{\beta}d\tau\sum_{\mathbf{q}}\mathcal{L}}~.\label{eq:GL_S}
\end{equation}
To calculate $F[\Delta]$, we perform an expansion $\Delta(\mathbf{k})=\Delta_{0}\delta_{\mathbf{k},\mathbf{0}}+\delta\Delta(\mathbf{k})$
around the extremum of $F[\Delta]$. Here, $\Delta_{0}$ represents the mean-field value at temperature $T$, while $\delta\Delta(\mathbf{k})$ represents the fluctuations of the order parameter.
By minimizing the GL free energy $\frac{\partial F[\Delta]}{\partial\Delta_{0}}=\frac{\partial F[\Delta]}{\partial\bar{\Delta}_{0}}=0$, the mean field order parameter $\Delta_{0}$ can be determined from the self-consistent
 gap equation 
\begin{equation}
1=\frac{1}{N}\sum_{\mathbf{q}}\frac{U|\Gamma(\mathbf{q},\mathbf{0})|^2}{2\epsilon(\mathbf{q})}\tanh\frac{\beta\epsilon(\mathbf{q})}{2}, \label{eq:self-eq}
\end{equation}
where $\epsilon(\mathbf{q})=\sqrt{\left| U\Gamma(\mathbf{q},\mathbf{0})\Delta_{0}\right|^{2}+\mu^{2}}$
denotes the energy dispersion of Bogoliubov quasiparticles. Eq.~\eqref{eq:self-eq} can be simplified in the presence
of time-reversal symmetry $u^{}_{-\mathbf{q},-}=u_{\mathbf{q},+}^{*}$ such that
$\Gamma(\mathbf{q}, \mathbf{0})=1$.  With $\Delta_{0}$ being a constant, one obtains $U\Delta_{0}/T_{\mathrm{MF}}=2$  at half-filling 
with $T_\mathrm{MF}$ as the mean-field critical temperature and $U\Delta_0$ as the pairing gap (see SM \cite{SM}),
which is larger than the ratio ($\sim1.7$) from a conventional BCS theory.

Going beyond mean field and including the fluctuations, we have $F[\Delta]=F_{0}+F_{2}+\mathcal{O}(|\delta\Delta|^{4})$
up to the second order of $\delta\Delta(\mathbf{k})$. In particular,
$F_{0}$ recovers the grand potential 
\begin{equation}
F_{0}=\sum_{\mathbf{k}}\left[U\vert\Delta_{0}\vert^{2}-2\ln(1+e^{-\beta\epsilon(\mathbf{k})})+\epsilon(\mathbf{k})\right].\label{eq:grandpot}
\end{equation}
The second order free energy $F_{2}\equiv\sum_{\mathbf{k}}\mathcal{L}[\delta\Delta]$
describes the Gaussian fluctuations with 
\begin{equation}
\mathcal{L}[\delta\Delta]=\left[U-U^{2}\chi(\mathbf{k})\right]\delta\bar{\Delta}(\mathbf{k})\delta\Delta(\mathbf{k})~,\label{eq:L_Delta}
\end{equation}
where $\chi(\mathbf{k})$ is the four-point correlation function,
\begin{align}
\chi(\mathbf{k})\equiv & \frac{T}{N}\sum_{q}\vert\Gamma(\mathbf{q},\mathbf{k})\vert^{2}[\mathcal{G}(q+k/2)\mathcal{G}(-q+k/2)\nonumber \\
 & +\mathcal{F}(q+k/2)\mathcal{F}(-q+k/2)]~,\label{eq:chi}
\end{align}
with Gor'kov's normal and anomalous Green's functions $\mathcal{G}(q)$
and $\mathcal{F}(q)$ ($q=(\mathbf{q},-i\omega_n)$) (see SM-I \cite{SM}). 
In contrast to conventional superconductors, the Bloch wavefunctions play a significant role in both the
effective interaction and the quasiparticle dispersion. The prefactor $|\Gamma(\mathbf{q},\mathbf{k})|^{2}$ 
in Eq.~(\ref{eq:chi}) highlights the importance of the form of the Bloch wavefunctions. The significance of $|\Gamma(\mathbf{q},\mathbf{k})|^{2}$ is that, the pairing strength of a finite momentum Cooper pair is weighed by $\Gamma(\mathbf{q},\mathbf{k})$, such that $\chi(\mathbf{k})$ is $\mathbf{k}$-dependent. This generates a finite superfluid weight even though the effective mass of electrons diverges for a completely flat band. As we show below, the form factor encodes the quantum metric effects. 

\emph{\color{blue}Superfluid weight, BKT transition and quantum metric.---} In general, the form factor can be expanded as a function of $\mathbf{k}$ up to the second order:
\begin{equation}
\vert\Gamma(\mathbf{q},\mathbf{k})\vert^{2}=\gamma_{0}(\mathbf{q})-\sum_{ab}g_{ab}(\mathbf{q})k_{a}k_{b},\label{eq:chi2expan}
\end{equation} 
where the absence of a linear term is due to the stability of the mean-field ansatz. 
For a time-reversal invariant system where $u_{\mathbf{q},+}=u_{\mathbf{-q},-}^{*}\equiv u_{\mathbf{q}}$,
$\gamma_{0}(\mathbf{q})$ becomes the inner product $\gamma_{0}(\mathbf{q})\equiv\vert\langle u_{\mathbf{q}}\vert u_{\mathbf{q}}\rangle\vert^{2}=1$
and $g_{ab}(\mathbf{q})$ is the Fubini-Study
metric \citep{1980CMaPh76289P,2010arXiv1012.1337C} with components
\begin{equation}
g_{ab}(\mathbf{q})\equiv\mathrm{Re}\langle\partial_{\mathbf{q}_{a}}u_{\mathbf{q}}\vert(1-|u_{\mathbf{q}}\rangle\langle u_{\mathbf{q}}|)\vert\partial_{\mathbf{q}_{b}}u_{\mathbf{q}}\rangle,\label{eq:gab}
\end{equation}
which measures the Bures distance between two quantum states. The quantum metric $g_{ab}(\mathbf{q})$ characterizes
how the Bloch states interfere with each other.
The appearance of the quantum metric
in Eq.~\eqref{eq:chi}  and Eq.~\eqref{eq:chi2expan} clearly illustrates the crucial
role of the quantum geometric effect in determining superconductivity
fluctuations. Importantly, this effect is not evident at the mean-field level, as demonstrated by Eq.~\eqref{eq:grandpot}.
After integrating out
the Matsubara frequency along with the expansion in Eq.~(\ref{eq:chi2expan}),
we have  
$\chi(\mathbf{k})=\chi_{0}-\frac{1}{8}\sum_{ab}\mathsf{\chi}_2^{ab}k_{a}k_{b}$,
with the explicit form as \vspace{-2pt}
\begin{align}
\chi_{0} & =\frac{1}{N}\sum_{\mathbf{q}}\frac{\gamma_{0}(\mathbf{q})}{2}\frac{1}{\epsilon(\mathbf{q})},\label{eq:chi0}\\
\chi_{2}^{ab} & =\frac{2U^{2}\Delta_{0}^{2}}{N}\sum_{\mathbf{q}}\frac{\tanh\left(\frac{\beta\epsilon(\mathbf{q})}{2}\right)}{\epsilon(\mathbf{q})}g_{ab}(\mathbf{q}).\label{eq:chiab}
\end{align}
Thus in the continuum limit $F_2= \int d \mathbf r \mathcal L[\delta\Delta]$, we reach an effective theory  $\mathcal L[\delta\Delta]$ 
\begin{equation}
\mathcal{L}[\delta\Delta]=\frac{1}{8}\sum_{ab}\mathsf{D}_{s}^{ab}\partial_{a}\delta\bar{\Delta}\partial_{b}\delta\Delta.\label{eq:LdeltaD}
\end{equation} 

The factor $\mathsf{D}_{s}^{ab}\equiv  \mathcal A_\mathrm{uc}^{-1} \chi_{2}^{ab}$, with $\mathcal A_\mathrm{uc}$ being the area of the unit cell,
depends on $g_{ab}(\mathbf{q})$. This indicates that the dynamics of the fluctuation of the order parameter is governed by the quantum metric.  Assuming the presence of spatial phase fluctuations
$\delta\Delta(\mathbf{r})=\Delta_{0}e^{2i\theta(\mathbf{r})}-\Delta_{0}\simeq2i\theta(\mathbf{r})\Delta_{0}$
(and ignoring the amplitude fluctuations), we obtain the effective Lagrangian 
\begin{equation} 
\mathcal{L}[\theta]=\frac{1}{2}\text{\ensuremath{\sum_{ab}}}\mathsf{D}_{s}^{ab}\partial_{a}^{}\theta\partial_{b}^{}\theta~,  \label{eq:D}
\end{equation} 
and the supercurrent $j_{b}^{}=\sum_{a}\mathsf{D}_{s}^{ab}\partial_{a}^{}\theta$. Here, $\mathsf{D}_{s}^{ab}$ is identified as the superfluid weight which is consistent with previous studies \cite{2015NatCo68944P,PhysRevB.95.024515,PhysRevB.106.014518}.

We can now determine the BKT transition temperature as $T_{\mathrm{BKT}}=\pi\sqrt{\mathrm{det}\mathsf{D}_{s}^{ab}}/8$ \cite{PhysRevLett.39.1201, PhysRevB.106.014518}.  For an isotropic superconductor with a flat Bogoliubov quasiparticle band, we obtain a simple relation 
$T_{\mathrm{BKT}} = \frac{\pi U\Delta_0}{8\mathcal A_\mathrm{uc}}\sqrt{\mathrm{det}(\bar{g}_{ab})}$ near the half-filling,
where the average of the quantum metric over the Brillouin zone is:
\begin{equation}
\bar{g}_{ab}= \frac{1}{N}\sum_{\mathbf q} g_{ab}(\mathbf{q}). 
\label{eq:aver_gamma}
\end{equation}
Interestingly, when $\bar{g}_{ab}$ is sufficiently large, one can show that $T_\mathrm{BKT}$ approaches $T_\mathrm{MF}$. This is very different from the BCS theory predictions. According to BCS theory, $\mathsf D_{s}^{ab} = \delta_{ab} n_s/m^*$  with $n_s$ being the superfluid density and $m^*$ is the effective mass. In that case, $T_{\mathrm{BKT}}=\pi n_s/(8m^*)$ goes to zero for flat band superconductors.

\emph{\color{blue} Upper critical field $H_{c2}$.---} 
Another important physical quantity of a superconductor is the (orbital) upper critical field which is expected to be infinite according to the BCS theory for flat band superconductors. As the mean-field order parameter is suppressed by vortex excitations around $H_{c2}$, we can derive the GL free energy from Eq.~\eqref{eq:GL_S}  by assuming a vanishing mean field, $\Delta_{0}=0$. An effective Lagrangian can be obtained after integrating out the fermion field, and for an isotropic system, we have $F_2= \int d \mathbf r \mathcal L[\delta\Delta]$, with
\begin{equation}\vspace{-2pt}
\mathcal{L}[\delta\Delta]\!=\!\frac{1}{2m^{*}}\vert\nabla\delta\Delta\vert^{2}+a(T)\vert\delta\Delta\vert^{2}+\mathcal{O}(\vert\delta\Delta\vert^{4}),\label{eq:Lhc2}
\end{equation}
and \vspace{-2pt}
\begin{align}
\frac{1}{2m^{*}} & =\frac{\beta U^{2}\bar{\nu}}{4\mathcal A_\mathrm{uc}}\sqrt{\mathrm{det}(\bar{g}_{ab})},  \label{eq:mass} \\
a(T) & =  \frac{ 4U - U^2 \beta\bar{\nu}\bar{\gamma}_{0}}{4\mathcal A_\mathrm{uc} }.
\end{align}
Here, $\bar{\gamma}_{0}=\frac{1}{N}\sum_{\mathbf{q}}\gamma_{0}(\mathbf{q})=1$ and $\bar{\nu}=\frac{2(1-2\nu)}{\ln(\nu^{-1}-1)}$ with $\nu$ being the filling factor.
It is interesting from Eq.~\eqref{eq:mass} that the quantum metric $\sqrt{\mathrm{det}(\bar{g}_{ab})}$ gives rise to a finite effective mass for the Cooper pairs.
The magnetic field can be included in the free energy by the minimal coupling $-i\nabla\rightarrow-i\nabla+2e\mathbf{A}$ in Eq.~(\ref{eq:Lhc2}). Then, the upper critical field $H_{c2}$ can be determined using the standard GL approach \cite{altland2010condensed} and we obtain:
\begin{equation}
H_{c2}=\frac{\Phi_0}{2\pi\sqrt{\det(\bar{g}_{ab})}}\frac{\vert T-T_\mathrm{MF}\vert}{T_\mathrm{MF}}, \label{eq:hc2}
\end{equation}
where $\Phi_0=h/2e$ is a flux quantum.
It is clear that a finite average quantum metric $\bar{g}_{ab}$ would give rise to a finite $H_{c2}$. 
Moreover, using the condition $H_{c2}\xi^{2}= \Phi_{0}/2\pi$, 
the superconducting coherence length can be written as:
\begin{equation}
\xi=\sqrt{\frac{T_\mathrm{MF}}{\vert T- T_\mathrm{MF}\vert}  }[\mathrm{det}(\bar{g}_{ab})]^{\frac{1}{4}}.
\label{eq:cohlen}
\end{equation}
As expected, the expression of $\xi$ is dramatically different from the BCS relation $\xi = \frac{\hbar v_{F}}{\Delta}$ which vanishes as $v_F$ goes to zero. It is important to note that at zero temperature, the coherence length is extracted as $\xi (T=0)=[\mathrm{det}(\bar{g}_{ab})]^{\frac{1}{4}}$ which is the size of optimally localized Wannier functions constructed from the Bloch states of the flat band \cite{PhysRevB.56.12847,2015NatCo68944P} (also see SM-II~\cite{SM}). This is a very interesting result as $\xi$ is shortest at $T=0$ and it is independent of the interaction strength $U$. This means that the minimal Cooper pair size is purely determined by the quantum metric at $T=0$ and a stronger interaction cannot reduce the Cooper pair size. When the temperature increases, the thermal energy competes with the interaction energy, which is incorporated into $T_\mathrm{MF}$, and $\xi$ depends on the interaction strengths at finite temperatures.

\emph{\color{blue} Application to TBG---} 
Recently, superconductivity has been observed in TBG with extremely small $v_F$, estimated to be around $1,000$m/s. In this section, we employ the GL theory developed above to explain the observed superconducting coherence length which is about 20 times larger than the one estimated from BCS theory.  As shown in Eq.~(\ref{eq:cohlen}), to calculate the coherence length $\xi$,
we use the Bloch states of the moir\'e bands of the Bistrizer-Macdonald model which incorporates both the quantum metric and the band dispersion of TBG near the magic angle \cite{2011PNAS..10812233B}. The details of the model is given in the SM-III~\cite{SM}.  In Fig.~\ref{Fig:TBG}(a), the band structure at twisted angle $\theta=1.08^{\circ}$ is shown and the bandwidth is extremely narrow which is in the order of $1$meV. For simplicity, only the moir\'e bands originated from one valley of the original BZ is shown.

It is clear that the hole (solid line) and the electron (dashed line) bands touch at the Dirac points at $\mathsf{K}$ and $\mathsf{K}^{\prime}$ of the moir\'e Brillouin zone. Here the chemical potential $\mu=0$ is set at the Dirac point. 
To model the superconducting phase, we assume an effective singlet pairing potential \cite{PhysRevB.101.060505,PhysRevLett.123.237002}  
$H_{\mathrm{int}}=-U\int d^2\mathbf r\sum_{\ell\xi}\psi_{\uparrow\rho,\ell\xi}^{\dagger}(\mathbf{r})\psi_{\downarrow\bar{\rho},\ell\xi}^{\dagger}(\mathbf{r})\psi_{\downarrow\bar{\rho},\ell\xi}(\mathbf{r})\psi_{\uparrow\rho,\ell\xi}(\mathbf{r})$
with the valley indices $\rho=\pm$, sublattices $\xi=A,B$, and layer index
$\ell$, while ignoring other correlation effects induced by interactions between the electrons \cite{2018Natur.556...80C,PhysRevLett.124.097601}.  
Here $\psi_{\sigma\rho,\ell\xi}$ denotes a Fermion operator in the continuum limit. 
It is important to note that the pairing does not need to be a singlet in TBG but we focus on the quantum metric effect here and ignore the possible complications due to other pairing symmetries.

We can determine the coherence length within the GL theory by taking into account both the quantum metric and the band dispersion. The detail calculations involving the band dispersions are given in the Supplemental Material (see SM-IV~\cite{SM}).  
In Fig.~\ref{Fig:TBG}(a), we plot the quantum metric as a function of momentum.
It is interesting to note that the quantum metric diverges around the two Dirac points. To avoid the divergence of the quantum metric in the calculations, we set a small energy gap at the Dirac point. For TBG, the total coherence length $\xi$ consists of both
the quantum metric contribution ($\xi_\mathrm{qm}$) and band dispersion contribution. The $\xi_\mathrm{qm}$ has the similar form to Eq.~\eqref{eq:cohlen} with the average quantum metric $\bar g_{ab}$ evaluated at 0.2K (see SM-IV~\cite{SM}).
 Fig.~\ref{Fig:TBG}(b) shows both the total $\xi$ and 
the quantum metric contribution $\xi_\mathrm{qm}$ 
at different chemical potentials $\mu$ which are relevant to the experimental regime.
In the regime from $\mu =-0.1$meV to $\mu =-0.2$meV, there is a decrease 
in $\xi$ when the chemical potential decreases. This is due to the decrease of the quantum metric contribution $\xi_\mathrm{qm}$.  When the chemical potential decreases further, there is an increase of $\xi$, due to the increase of the quantum metric contribution and and the Fermi velocity contribution.

Amazingly, similar $\xi$ dependence on the chemical potential was observed in the experiment \cite{2023Natur.614..440T}. Without fine tuning of parameters, we obtained $\xi \approx 30 \mathrm{nm}$ at $0.2$K as shown in Fig.~\ref{Fig:TBG}, which is comparable to the experimental values of about $55 \mathrm{nm}$. The estimated $H_{c2}$ is also plotted in Fig.~\ref{Fig:TBG}(b). 
The highest $H_{c2}\sim 0.27\mathrm{T}$ which is also comparable to the experimental value of $0.1\mathrm{T}$. 

\emph{\color{blue} Conclusion.---} 
We developed a GL theory for flat band superconductors which includes the quantum metric effects. The GL theory allows us to derive many of the important physical quantities of superconductors in terms of the quantum metric of the Bloch electrons as summarized in Table~\ref{tab:summary}. 
Importantly, we found that the coherence length, which is expected to be zero from a conventional BCS theory, is finite for flat bands with quantum metric. 
Physically, the size of the optimally localized Wannier functions, which is governed by the quantum metric, determines the superconducting coherence length at zero temperature. 
Hence, the quantum metric sets a fundamental length scale in flat band superconductors.
By calculating the quantum metric of TBG, we explained qualitatively the dependence of the coherence length on the chemical potential in the experiment. The  GL theory developed in this work provides a general framework to understand the unconventional properties of flat band superconductors.

\emph{\color{blue} Acknowledgements ---}
We thank Jeanie Lau for informing us about their experimental results before publication. We acknowledge valuable discussions with Tai-Kai Ng, Adrian Po, W. Huang and Y.-B. Yang. K.T.L. acknowledges the support of the Ministry of Science and Technology, China, and the Hong Kong Research Grants Council through Grants No. 2020YFA0309600, No. RFS2021-6S03, No. C6025-19G, No. AoE/P-701/20, No. 16310520, No. 16310219, No. 16307622, and No. 16309718.

%

\end{document}